\documentclass{PoS}
\newcommand{\MG} {\textsc{mg5}\_a\textsc{mc}}

\newcommand*{\tmax}{transMAX}
\newcommand*{\tmin}{transMIN}

\newcommand*{\ttbar}{$\mathrm{t}\overline{\mathrm{t}}$~}

\newcommand*{\alpS}{$\alpha_S$}
\newcommand*{\alpISR}{$\alpha_S^{ISR}$}
\newcommand*{\alpFSR}{$\alpha_S^{FSR}$}
\newcommand*{\SHERPA} {{\textsc{sherpa}}~}
\newcommand*{\PYTHIA} {{\textsc{pythia}}~}
\newcommand*{\POWHEG} {{\textsc{powheg}}~}
\newcommand*{\HERWIG} {{\textsc{herwig}}~}
\newcommand*{\DIRE} {{\textsc{dire}}~}
\newcommand*{\pt}{\ensuremath{p_{\mathrm{T}}}~}
\newcommand{\CP}{\textsc{cuetp8m2t4}}
\newcommand{\CO}{\textsc{cuetp8m1}}

\title{Quantum Chromodynamics Monte Carlo Tuning Studies in CMS}

\ShortTitle{QCD Monte Carlo Tuning}

\author{\speaker{Efe Yazgan}\thanks{On behalf of the CMS Collaboration}\\
        Institute of High Energy Physics, Chinese Academy of Sciences, Beijing, China\\
        E-mail: \email{efe.yazgan@cern.ch}}

\abstract{Recent QCD Monte Carlo tuning studies done in the CMS Collaboration are presented. Jet kinematics, jet substructure, and underlying event measurements in top quark pair events are discussed. New CMS PYTHIA 8 event tunes are presented, exploiting Monte Carlo configurations with consistent parton distribution functions and strong coupling parameter values in the matrix element and the parton shower, at leading order (LO), next-to-leading order (NLO) and next-to-next-to-leading order (NNLO). Predictions from \PYTHIA8 obtained with tunes based on NLO or NNLO PDFs are shown to reliably describe minimum-bias and underlying-event data with a similar level of agreement to predictions from tunes using LO PDF sets. The tunes are validated with a range of different measurements and matrix element-parton shower merged configurations.}

\FullConference{XXVII International Workshop on Deep-Inelastic Scattering and Related Subjects - DIS2019\\
		8-12 April, 2019\\
		Torino, Italy}

\begin{document}

\section{Introduction}
Hadron collisions at particle colliders are modelled using quantum chromodynamics (QCD) Monte Carlo (MC) simulation codes. These codes consist of several parts. 
The hard scattering is modelled by simulating particles from the hadronization of partons kinematics of which are calculated using perturbation theory. 
Partons from initial-state radiation (ISR) and final-state radiation (FSR) are modelled by a parton shower (PS) algorithm. 
The underlying event (UE), a non-perturbative component, refers to the activity resulting from additional interactions of partons in the hadron collision. 
It consists of beam-beam remnants (BBR), and the particles from the additional relatively softer parton-parton scatterings referred to as the multiple-parton interactions (MPI). 
To obtain a reliable description of the UE, adjustable parameters in standard MC event generators, such as \PYTHIA8~\cite{pythia8}, \HERWIG\cite{Bellm:2015jjp}, and \SHERPA\cite{Gleisberg:2008ta}, need to be tuned to the data. 
Another non-perturbative ingredient is the parton distribution function (PDF) used both in the hard partonic matrix element (ME) calculation, the PS and the MPI model. \alpS($M_Z$) is a free parameter and its value used in simulations at LO or NLO are typically different. Different strategies are adopted in different collaborations; CMS~\cite{Khachatryan:2015pea} and ATLAS~\cite{ATL-PHYS-PUB-2014-021} tunes are traditionally based on LO PDFs, \PYTHIA8~\cite{Skands:2014pea} tunes are mostly based on LO PDFs, new tunes are based on NNLO PDFs, and \HERWIG7~\cite{Gieseke:2016fpz} on NLO PDFs. 
Merging schemes, such as the $MLM$~\cite{mlm}, allow the combination of predictions of jet production using ME calculations with those from PS emissions for soft and collinear parton radiation at leading-log (LL) accuracy without double counting or dead regions. Merged calculations capture some higher-order corrections with respect to the formal order of the ME calculation. 
Using the same PDF set and \alpS~value in the PS-ME merged calculations in the simulation of the various components is advocated in Ref.~\cite{Cooper:2011gk}, and by the \HERWIG7 and \SHERPA Collaborations. In the \PYTHIA8 tunes produced prior to Ref. \cite{Sirunyan:2019dfx}, \alpS($M_Z$) values in general are not selected to be the same as those used in the PDFs. For example, in the Monash tune, the \alpFSR($M_Z$), set to 0.1365, is obtained by fitting standalone-\PYTHIA8 predictions to LEP event-shape measurements, \alpISR($M_Z$) is assumed to be equal to \alpFSR($M_Z$), and the \alpS$(M_Z)$ values in hard scattering and MPI are set to 0.130~\cite{Skands:2014pea}. 
In this note, we discuss the preferred values of \alpISR($M_Z$) and \alpFSR($M_Z$) obtained from jet multiplicity, substructure and UE event measurements in \ttbar events and discuss new CMS \cite{CMS} tunes based on consistent PDF and \alpS($M_Z$) values.  
 
\section{Revisiting Parton Shower Parameters and Tunes}
The previous default CMS \PYTHIA8 tune, CUETP8M1~\cite{Khachatryan:2015pea}, does not describe the central values of $\sqrt{s}=$ 13 TeV UE data very well~\cite{Sirunyan:2019dfx} or the UE evolution from 1.96 to 7 TeV in UE in Z+jets~\cite{Sirunyan:2017vio}. It is also observed that the predictions of the POWHEG generator \cite{powhegv23} interfaced with the \PYTHIA8 code with the CUETP8M1 tune overshoot the data for large multiplicities where jets are produced by the PS ~\cite{CMS:2016kle} for both $\sqrt{s}=$ 8 and 13 TeV data. 

\subsection{Jet Multiplicity in \ttbar events and \alpISR}
Tuning \alpISR~using 8 TeV jet multiplicity in \ttbar events using the PS dominated region yields \alpISR = 0.1108$^{+0.0145}_{-0.0142}$~\cite{CMS:2016kle}. The tuned \alpISR($M_Z$) value agrees with the PDG value of \alpISR($M_Z$) = $0.1181\pm0.0011$ \cite{PhysRevD.98.030001} well within uncertainties. It is found that to describe the number of jet distribution in \ttbar events, the \PYTHIA8 ISR parameter {\it Rapidity Ordering} needs to be used ~\cite{Sirunyan:2019dfx}. 
Rapidity ordering imposes an additional constraint on the \pt-ordered emissions reducing the parton emission phase space. 
 The parton emission probability is mainly constrained by the jet activity and the interplay between the hard and soft parts of the parton emissions.~However, it does not strongly constrain the UE. Therefore, \alpISR($M_Z$) constrained by \ttbar jet kinematics is used as a fixed input parameter in deriving a new UE tune, \CP~\cite{CMS:2016kle}. This new tune, with a lower \alpS$^{ISR}(M_Z)$, with respect to the CUETP8M1 tunes is found to improve the description of \ttbar kinematics, as as the overall description of observables at $\sqrt{s}=8$ and 13 TeV. 

\subsection{Jet substructure in \ttbar events and \alpFSR($M_Z$)}
Jet substructure observables are measured using \ttbar events in the lepton+jets channel at $\sqrt{s}=13$ TeV \cite{PhysRevD.98.092014} using the \CP~tune for the \ttbar signal and \CO tune for other simulated samples. Jets are studied with different substructure observables using charged and neutral particles and charged-particle-only jet constituents. Inclusive jet samples as well as samples enriched in bottom, light-quark, or gluon jets are selected. 
Substructure observables at particle level are compared to NLO predictions from \POWHEG+\PYTHIA8, \HERWIG7, \SHERPA and \DIRE2 \cite{Hoche:2015sya}. 
Predictions with the default tunes do not yield a good overall description of the data, in particular, for particle multiplicity-related observables. 
These measurements help tune models. As a first step in this direction, the angle between the groomed subjets is used to extract \alpFSR=0.115$^{+0.015}_{-0.013}$ at LO+LL accuracy and using CMW rescaling \cite{Catani:1990rr}. This value agrees with the PDG value of \alpISR($M_Z$)  well within uncertainties. A more precise \alpFSR($M_Z$) could be obtained once calculations of top-quark decays with multiple emissions at higher order with PS at approximate NLL become available. These measurements also provide the data to improve analytical QCD calculations with higher-order corrections, for infrared-and/or collinear-safe observables.

\subsection{Underlying event in \ttbar events and \alpFSR($M_Z$)}
UE activity in \ttbar dilepton events are measured by CMS at $\sqrt{s}=13$ TeV.
The measurements are based in particle-flow reconstruction. 
The UE contribution is isolated by removing charged particles associated with the decay products of the \ttbar event candidates as well as with pileup interactions from the set of reconstructed charged particles for each event.
The observables and categories chosen for the measurements enhance the sensitivity to \ttbar modeling, MPI, color reconnection (CR) and the choice of \alpS($M_Z$) in the \PYTHIA8. 

Most of the comparisons indicate a fair agreement between the data and the \POWHEG + \PYTHIA8 setup with the \CP tune~\cite{CMS:2016kle}, 
but disfavor the setups in which MPI and CR are switched off.
The data also disfavor the default configurations in \POWHEG+\HERWIG++, \POWHEG+\HERWIG7, and \SHERPA.
It has been furthermore verified that, as expected, the choice of the next-to-leading-order matrix-element generator does not impact significantly the expected characteristics of the UE by comparing predictions from \POWHEG and \MG~\cite{Alwall:2014hca}, both interfaced with \PYTHIA8.

The UE measurements in \ttbar events test the hypothesis of universality of UE at an energy scale two times the top quark mass which is considerably higher than the ones at which UE models have been studied in detail.
The results also show that a value of \alpFSR$(M_{Z})=0.120\pm0.006$
is consistent with the data and the corresponding uncertainties
translate to a variation of the renormalization scale by a factor of $\sqrt{2}$ that has already been adopted in CMS \ttbar measurements. 

\subsection{New CMS UE tunes}
A set of new 13 TeV \PYTHIA8.226 UE tunes is obtained with different choices of values of \alpS$(M_Z)$ used in the modeling of the ISR, FSR, hard scattering, and MPI, as well as the order of its evolution as a function of the four-momentum squared $Q^2$. We distinguish the new tunes according to the order of the NNPDF3.1 PDF set~\cite{Ball:2017nwa} used: LO-PDF, NLO-PDF, or NNLO-PDF. The tunes are labeled as CPi, where i=1,2,...,5. CP1 and CP2 are based on LO, CP3 on NLO, and CP4 and CP5 on NNLO NNPDF3.1 set. 
We fit charged-particle and \pt$^{sum}$~densities, measured in \tmin\ and \tmax\ regions as a function of \pt$^{max}$, as well as the charged-particle multiplicity as a function of pseudorapidity $\eta$, measured by CMS at $\sqrt{s}=13$ TeV~\cite{CMS:2015zev,Khachatryan:2015jna}. In addition, we also use the charged-particle and \pt$^{sum}$ densities as a function of the leading charged-particle \pt, measured in \tmin\ and \tmax\ by CMS at $\sqrt{s}=7$ TeV~\cite{CMS:2012zxa} and by CDF at $\sqrt{s}=1.96$ TeV~\cite{Aaltonen:2015aoa}. 

The parameters related to the simulation of the hadronization and beam remnants are not varied in the fits and are kept fixed to the values of the Monash tune.
Only five parameters related to the simulation of MPI, to the overlap matter distribution function~\cite{Sjostrand:1987su}, and to the amount of CR are constrained for the new CMS tunes. In all tunes, we use the MPI-based CR model~\cite{Sjostrand:2004pf}. 
The overlap distribution between the two colliding protons is modeled according to a double-Gaussian functional form with the parameters \texttt{coreRadius} and \texttt{coreFraction}. 

In the LO NNPDF3.1 set, \alpS($M_Z$) = 0.130, whereas for the NLO and NNLO NNPDF3.1 sets, \alpS($M_Z$) = 0.118. 
Irrespective of the specific PDF used, predictions from the new tunes reproduce well the UE measurements at center-of-mass energies $\sqrt{s}=1.96$ and 7 TeV. A significant improvement in the description of UE measurements at 13 TeV is observed with respect to predictions from the previous tunes that were extracted using data at lower collision energies.
For the first time, predictions based on higher-order PDF sets are shown to give a reliable description of minimum-bias (MB) and UE measurements, with a similar level of agreement as predictions from tunes using LO PDF sets. Predictions of the new tunes agree well with the data for MB observables measured in the central ($|\eta|< 2.4$) and forward ($3.2 <|\eta|< 4.7$) regions.
The CP tunes simultaneously describe the number of charged particles produced in diffractive processes and MB collisions.
Neither the CP tunes nor the CUETP8M1 tune describe the very forward region ($-6.6<\eta<-5.2$) well. Measurements sensitive to double-parton scattering contributions are reproduced better by predictions using the LO PDF set in the UE simulation, without ISR rapidity ordering. 

The UE simulation provided by the new tunes can be interfaced to higher-order and multileg matrix element generators, such as \POWHEG and \MG, without degrading the good description of UE observables. Such predictions also reproduce well observables measured in multijet final states, Drell--Yan, and top quark production processes. The central values of the normalized \ttbar cross section in bins of the number of additional jets predicted by \POWHEG~+\PYTHIA8 overestimate the data when a high value of \alpISR($M_Z$)$\sim0.13$ is used (CMS \PYTHIA8 CP1 and CP2 tunes). Even when \alpISR($M_Z$)=0.118 is used, the CP4 tune overestimates the data at high jet multiplicities. This is cured by the ISR rapidity ordering in the CP5 tune (as in the CUETP8M2T4 tune). Measurements of azimuthal dijet correlations are also better described when a value of \alpISR($M_Z$)=0.118 is used in predictions obtained with \POWHEG merged with \PYTHIA8. All details about the CP tunes can be found in \cite{Sirunyan:2019dfx}.

\section{Conclusions}
Monte Carlo event modeling is studied using jet multiplicity, jet substructure, and underlying event measurements in top quark pair events. 
These studies showed that the data is consistent with a lower value of strong coupling parameter for both initial and final state radiation in the parton shower when merged configurations are used. 
New CMS \PYTHIA8 multipurpose tunes, aiming for a consistent description of UE and MB observables at several collision energies and a reliable prediction of the UE simulation in various processes when merged with higher-order ME calculations are presented. 
These new tunes exploit Monte Carlo configurations with consistent parton distribution functions and strong coupling parameter values in the matrix element and the parton shower, at leading order (LO), next-to-leading order (NLO) and next-to-next-to-leading order (NNLO). 
Predictions from \PYTHIA8 obtained with tunes based on NLO or NNLO PDFs are shown to reliably describe minimum-bias and underlying-event data with a similar level of agreement to predictions from tunes using LO PDF sets along with a wide range of different measurements.

\end{document}